\documentclass[floatfix,footinbib,reprint,twocolumn,aip,preprintnumbers,amsmath,amssymb]{revtex4-1}

\usepackage{graphics}

\usepackage{subfigure}
\usepackage{enumitem}  
\usepackage{bbold}
\DeclareGraphicsExtensions{.png,.pdf}
\usepackage{epsfig}
\usepackage{todonotes}
\usepackage{bm}
\usepackage{color}
\usepackage{makeidx}
\usepackage{hyperref}

\usepackage[T1]{fontenc}
	
\usepackage{booktabs}
\AtBeginDocument{
\heavyrulewidth=.08em
\lightrulewidth=.05em 
\cmidrulewidth=.01em
\belowrulesep=.65ex
\belowbottomsep=0pt
\aboverulesep=.4ex
\abovetopsep=0pt
\cmidrulesep=\doublerulesep
\cmidrulekern=.5em
\defaultaddspace=.5em
}
\usepackage{multirow}

\usepackage[utf8]{inputenc}
\usepackage{xcolor}

\begin{document}
\title{\bf Caloric effects around phase transitions in magnetic materials described by ab initio theory: The electronic glue and fluctuating local moments}

\author{Eduardo Mendive-Tapia}
\affiliation{Department of Computational Materials Design, Max-Planck-Institut für Eisenforschung GmbH, 40237 Düsseldorf, Germany}
\affiliation{Department of Physics, University of Warwick, Coventry CV4 7AL, U.K.}

\author{Julie B. Staunton*}
\affiliation{Department of Physics, University of Warwick, Coventry CV4 7AL, U.K.}

\date{\today}

\begin{abstract}
We describe magneto-, baro- and elastocaloric effects (MCEs, BCEs and eCEs) in materials which possess both discontinuous (first-order) and continuous (second-order) magnetic phase transitions. Our {\it ab initio} theory of the interacting electrons of materials in terms of disordered local moments (DLMs) has produced explicit mechanisms for the drivers of these transitions and here we study associated caloric effects in three case studies where both types of transition are evident. Our earlier work had described FeRh's magnetic phase diagram and large MCE. Here we present calculations of its substantial BCE and eCE.  We describe the MCE of dysprosium and find very good agreement with experimental values for isothermal entropy ($\Delta S_{iso}$) and adiabatic temperature ($\Delta T_{ad}$) changes over a large temperature span and different applied magnetic field values. We examine the conditions for optimal values of both $\Delta S_{iso}$ and $\Delta T_{ad}$ that comply with a Clausius-Clapeyron analysis, which we use to propose a promising elastocaloric cooling cycle arising from the unusual dependence of the entropy on temperature and biaxial strain found in our third case study - the Mn$_3$GaN antiperovskite. We explain how both $\Delta S_{iso}$ and $\Delta T_{ad}$ can be kept large by exploiting the complex tensile strain-temperature magnetic phase diagram which we had earlier predicted for this material and also propose that hysteresis effects will be absent from half the caloric cycle. This rich and complex behavior stems from the frustrated nature of the interactions among the Mn local moments.
\end{abstract}
\maketitle

\section{Introduction}
\label{Intro}

The extent and state of magnetic order can change dramatically when a material is subjected to an external stimulus such as a magnetic field, applied pressure or strain. Magneto- and mechano-(baro- and elasto-)caloric effects that accompany these changes underpin the design of cooling cycles and attractive forms of solid state cooling~\cite{0034-4885-68-6-R04,SANDEMAN2012566,moya2014caloric}.
At a fundamental level the versatility of magnetic materials for such applications comes from the complex glue of electrons which both binds the nuclei of a material together and generates the magnetism. The spins of the septillions of interacting electrons cause atomic-scale, relatively long-lived magnetic moments to emerge from the electronic fluid and the moments’ behavior fixes properties such as the overall magnetization along with its resilence and response to magnetic fields. At high temperatures the moments disorder so that they average out to produce a low or zero magnetisation. At lower temperatures ordered patterns of the moments form changing the material’s properties. 

Over the last few years we have developed {\it ab initio} computational modelling which accurately accounts for this physics~\cite{PhysRevB.99.144424,PhysRevMaterials.3.101401}.  It describes how intrinsic magnetic properties change as the temperature is varied, a magnetic field applied, the material is compressed or strained etc. We can model both smooth (second-order) and sharp (first-order) transitions between different magnetic moment ordered patterns in multicomponent materials and calculate quantities to gauge usefulness for multicalorics and energy-efficient cooling applications. Our theory models the material at the sub-nanoscale, where the behavior of the many interacting electrons is carefully accounted for, to produce interactions among the magnetic moments. A statistical mechanical treatment describes quantitatively and directly the caloric properties such as the changes in entropy and temperature that happen when a magnetic field and/or mechanical stress is applied.

Our recent applications of the theory include the first-order ferromagnetic (FM) to antiferromagnetic (AFM) transition around room temperature and large magnetocaloric effect in the much studied FeRh ordered alloy~\cite{PhysRevB.89.054427}, metamagnetic critical fields in CoMnSi-based alloys~\cite{PhysRevB.87.060404}, FM, AFM and canted magnetic phases in lanthanide intermetallics~\cite{PhysRevLett.115.207201}, the magnetic field and temperature induced transitions between  long period helical AFM, fan and FM phases in the heavy lanthanide elements~\cite{PhysRevLett.118.197202}, the frustrated magnetism and mechanocaloric effects in the Mn-based antiperovskite nitrides~\cite{PhysRevB.95.184438}, temperature dependent permanent magnetic properties~\cite{PhysRevMaterials.3.101401}, and transitions between paramagnetic, ferrimagnetic, collinear AFM and non-collinear triangular AFM phases in the Mn$_3$A class of materials together with the influence of strain and volume change~\cite{PhysRevB.99.144424}.

In this paper we outline the underlying theory and computational approach (Sect.\ \ref{theory}). We illustrate the theory with a description of mechanocaloric effects in FeRh in the vicinity of its FM-to-AFM first order phase transition when subjected to pressure and/or strain (Sect.\ \ref{FeRh}). We then present results which show both the complexity and potential for exploitation for solid state cooling of the caloric effects that can occur in a material with a magnetic phase diagram containing both first- and second-order phase transitions. We discuss the implications on these aspects within the context of our modelling of the magnetocaloric effect in the heavy rare earth elements (Sect.\ \ref{HRE}). These examples serve as an introduction to our proposals for new cooling cycles which we illustrate by a further case study based on the elastocaloric effects of the antiperovskite nitride Mn$_3$GaN (Sect.\ \ref{Antiperovs}). We conclude with an outlook for the role of {\it ab initio} predictive modelling in the  multicalorics field.

\section{Temperature-dependent magnetic properties from first principles theory.}
\label{theory}

The {\it ab initio} modelling is based on the premise that important magnetic fluctuations in a material can be modelled as `local moments', a picture captured by a generalization of Density Functional Theory (DFT) for noncollinear spin polarization. For itinerant electrons a separation of time scales between fast and slow electronic degrees of freedom causes local moments with slowly varying orientations, $\{\hat{\bf{e}}_n\}$, to emerge from the interacting electrons of a  material with N atomic sites at positions $\{ {\bf R}_n \}$~\cite{0305-4608-15-6-018,PhysRevLett.69.371,PhysRevB.89.054427,PhysRevB.99.144424,PhysRevMaterials.3.101401}. This means that 'disordered local moments' (DLMs) are sustained by and influence the faster electronic motions. Their interactions with each other depend on the type and extent of the long-range magnetic order through the associated spin-polarized electronic structure~\cite{PhysRevB.89.054427} which itself adapts to the extent of magnetic order. For materials with rare earth lanthanide (RE) components the strongly correlated 4f electrons are treated with our self interaction correction (SIC) approach which is parameter free and incorporates Hund's rules naturally~\cite{PhysRevB.97.224415}. The crucial RE contribution to the magnetic anisotropy is accounted for using crystal field theory, calculating the crystal field coefficients within DFT using a robust numerical method~\cite{Patrick_2019}.

Ensemble averages over all the appropriately weighted noncollinear local moment orientational configurations are required for a realistic evaluation of the system’s magnetic properties. A set of magnetic order parameters, $\{\textbf{m}_n= \langle \hat{\bf{e}}_n \rangle \}$, specify the type and extent of magnetic order. 
An {\it ab initio} Gibbs free energy is given by the following expression:
\begin{equation}
\begin{split}
\mathcal{G}_1=-TS & + \Omega \big(\{\textbf{m}_n\},\textbf{H}, P, \sigma_{\alpha\beta}, T \big) \\
 & +\mathcal{G}_{RE}\big(\textbf{B}_{eff}(\{\textbf{m}_n\}),\textbf{H},P, \sigma_{\alpha\beta},  T \big)
\label{Eq1}
\end{split}
\end{equation}
 where $\Omega\big(\{\textbf{m}_n\},\textbf{H}, P, \sigma_{\alpha\beta}, T \big)$ is a magnetic energy of the material, which can include the effect of an external magnetic field $\textbf{H}$, applied pressure $P$ and mechanical stress $\sigma_{\alpha\beta}$. $\Omega \big(\{\textbf{m}_n\},\textbf{H}, P, \sigma_{\alpha\beta}, T \big)$ is obtained as an average over local moment configurations of the grand potential of the itinerant, interacting electrons of a material with spin polarization constrained to $\{\hat{\bf{e}}_n\}$~\cite{0305-4608-15-6-018,PhysRevLett.69.371,PhysRevB.89.054427,PhysRevB.99.144424,PhysRevMaterials.3.101401}. $S=\sum_n S_n$ is the total entropy of the local moments and $\mathcal{G}_{RE}\big(\textbf{B}_{ex}(\{\textbf{m}_n\}),\textbf{H}, P, \sigma_{\alpha\beta},T \big)$ describes the free energy of localized RE-4f electrons determined by an atomic-like Hamiltonian which includes the effective magnetic field $\textbf{B}_{ex}(\{\textbf{m}_n\})$ from the other electrons in the system and the crystal field~\cite{PhysRevMaterials.3.101401,Patrick_2019}. The equilibrium state of the system for specific values of the temperature, $T$, and applied field $\textbf{H}$ and/or $P$ and $\sigma_{\alpha\beta}$, is given by the set of order parameters $\{\textbf{m}_n\}$ which minimizes the Gibbs free energy function $\mathcal{G}_1$~\cite{PhysRevB.89.054427,PhysRevB.99.144424}.

The magnetic energy can be expressed in the form
\begin{widetext}
\begin{equation}
\Omega \big(\{\textbf{m}_n\},\textbf{H}, P, \sigma_{\alpha\beta}, T \big)=\Omega_0+f^{(2)}\big(\{\textbf{m}_n\} \big)+ \sum_{a>2}f^{(a)}\big(\{\textbf{m}_n\}\big)-\textbf{H}\cdot\sum_n\mu_n\textbf{m}_n
+\frac{1}{2}C_{\alpha\beta\gamma\kappa}\bm{\varepsilon}_{\alpha\beta}\bm{\varepsilon}_{\gamma\kappa}+\sigma_{\alpha\beta}\bm{\varepsilon}_{\alpha\beta},
\label{Eq2}
\end{equation}
\end{widetext}
where $\Omega_0$ is a constant, $\mu_n$ is the size of a local moment on a site $n$ and $\{f^{(a)}\}$ are order $a$ functions of $\{\textbf{m}_n\}$ which can depend on the lattice deformation, described by the strain tensor $\bm{\varepsilon}_{\alpha\beta}$ and the energy cost of mechanical stress given by the elastic moduli $C_{\alpha\beta\gamma\kappa}$ term.  The $f^{(2)}$ describe pairwise correlations between local moments and are obtained rigorously from an analysis of the linear response of the paramagnetic state to small applied fields~\cite{PhysRevB.99.144424}. They identify the dominant pair interactions between moments and the potentially most stable magnetic phases in a material, which can include non-collinear and long-period states in complex multi-atom unit cells. The $f^{(a>2)}$ describe higher order effects and produce a picture of effective `multi-site' magnetic interactions depending on how the electronic structure evolves with the state and extent of magnetic order. They are extracted from a linear regression fitting procedure to data produced by a large number of calculations for prescribed magnetic orders  $\{\textbf{m}_n\}$~\cite{PhysRevB.99.144424,PhysRevB.97.224415,PhysRevLett.118.197202,PhysRevB.89.054427,PhysRevMaterials.3.101401,PhysRevB.95.184438} and which exploits the symmetries of the feasible magnetic states.

In general, a caloric effect is quantified by the isothermal entropy change, $\Delta S_{iso}$, and the adiabatic temperature change, $\Delta T_{ad}$, induced in the thermodynamic conjugate of the external field applied and/or removed.
Our DFT-DLM theory can directly provide the total entropy as a function of the state of magnetic order, as well as the dependence on temperature, applied stimuli and crystal structure. It predicts the entropy changes formulated as a sum of three components. The first two are natural outputs from the DLM picture and are calculated directly. There is the contribution from the orientational disorder of the local moments, $\Delta S_{mag}=\Delta(\sum_n S_n)$, ~\cite{0953-8984-26-27-274210, PhysRevB.99.144424} and then an electronic piece estimated from the entropy change from alterations in the vicinity of the Fermi energy $E_{F}$ of the electronic density of states (DOS), i.e. $\frac{\pi^2}{3}k_{B}^2T \,n(\{\textbf{m}_n\},E_{F})$ where $n(\{\textbf{m}_n\},E_{F})$ is the DOS at the Fermi energy in a magnetic state specified by order parameters $\{\textbf{m}_n\}$.  The third component contains the lattice vibrational entropy and is currently estimated from a simple Debye model relying on $\theta_{D}$ the Debye temperature, which can be obtained from experiment or other first principles sources~\cite{CHEN2001947}.

\section{Baro- and elastocaloric effects in iron-rhodium.}
\label{FeRh}

For roughly stoichiometric compositions Fe$_{50}$Rh$_{50}$ orders into a paramagnetic (PM) B2 (CsCl) alloy phase. On cooling the alloy undergoes a second-order transition into a ferromagnetically ordered phase at a $T_c$ around 700K followed by its famous first-order ferromagnetic-to-antiferromagnetic (FM-AFM) transition at $T_t$. This FM-AFM magnetic transformation occurs near room temperature and shows one of the largest recorded magnetocaloric effects (MCEs) around $T_t$~\cite{NIKITIN1992234,doi:10.1063/1.360955} which deteriorates on subsequent magnetic and thermal cycling. The FM-AFM transition has been extensively studied, e.g.~\cite{Fallot,doi:10.1063/1.360955,doi:10.1063/1.369224,Kobayashi_2001,PhysRevB.72.214432,PhysRevB.77.184401,doi:10.1063/1.3556754,de_Vries_2013} and has been shown to be profoundly composition dependent. It vanishes in alloys with as little as a 2\% iron excess or deficiency and $T_t$ varies strongly with sample preparation, i.e.\ the extent of the long range B2 order. Building on several DFT studies~\cite{PhysRevB.46.2864,PhysRevB.67.064415,PhysRevB.83.174408,doi:10.1080/01411590412331316591,PhysRevB.85.174431} we applied the theory described above and were able to model the salient features of the alloy's behavior and obtain good agreement with experiment. We described {\it ab initio} the magnetic phase diagram of FeRh, its MCE and the hypersensitivity of the FM-AFM transition to compositional variation~\cite{PhysRevB.89.054427}.

The theory modelling produces a picture of local moments forming on the cubic sublattice occupied by Fe atoms which interact antiferromagnetically with each other. There is a competing energy benefit however if they align ferromagnetically so that the electronic density around the Rh sites on the other interpenetrating cubic sub-lattice becomes spin-polarized. A small number of Fe atoms replacing Rh ones greatly strengthen this ferromagnetic effect. It is this exquisite balance between AFM and FM mechanisms and their robust electronic origins which are at the centre of the compositional hypersensitivity and large negative MCE at the AFM-FM discontinuous transition~\cite{PhysRevB.89.054427}.  Here we report that the competition between the  FM and AFM interactions are also affected by the separations between Fe nearest neighbors and lead to pronounced barocaloric and elastocaloric effects around the FM-AFM discontinuous transition. We point out however that the modelling shows that magnetostructural effects are not the principal drivers of the transition but instead are consequences of it. 

\begin{table*}
\begin{center}
\begin{tabular}{ c | cc | cc }
 \hline\hline
 & $\Delta S_{iso}^{exp}$ (Jkg$^{-1}$K$^{-1}$) & $\Delta S_{iso}^{theo}$ (Jkg$^{-1}$K$^{-1}$) & $\Delta T_{ad}^{exp}$ (K) & $\Delta T_{ad}^{theo}$ (K)  \\
 \hline
MCE ($H=0\rightarrow$2T)           & between 12 and 20 & +22 & -7    & -7   \\
BCE ($P=0\rightarrow$1kbar)        & -12 & -23 & between 0 and +10  & +3 \\
eCE (strain$=0\rightarrow$ 0.05\%) & -8  & -23 & +3      & +5 \\
 \hline\hline
\end{tabular}
\caption{The MCE, BCE and eCE (compressive) of FeRh from the theory and comparison with experimental values (MCE~\cite{PhysRevLett.109.255901,SANDEMAN2012566,NIKITIN1992234,doi:10.1063/1.360955,PhysRevB.89.214105}, BCE~\cite{PhysRevB.89.214105} and eCE~\cite{PhysRevMaterials.2.084413}).}
\label{TableFeRh}
\end{center}
\end{table*}

The antiferromagnetic interactions between neighboring Fe atoms strengthen when their separations decrease as found by other authors~\cite{PhysRevB.46.2864}. We model the effect of pressure and compressive biaxial strain on the transition by assuming a Poisson ratio of $\frac{1}{3}$ and bulk and Young moduli both equal to 1800kbar as reasonable empirical values. The description of lattice fluctuations uses a Debye temperature of 400K. We estimate a spontaneous volume reduction of 0.7\% at the FM-AFM transition and calculate the variation of T$_t$ with pressure to be 3K/kbar.  Table~\ref{TableFeRh} summarises our results for the magnetocaloric, barocaloric (BCE) and elastocaloric (eCE) effects in iron rhodium together with a comparison with experimental data~\cite{PhysRevLett.109.255901,SANDEMAN2012566,NIKITIN1992234,doi:10.1063/1.360955,PhysRevB.89.214105,PhysRevMaterials.2.084413}. The MCE is calculated for an application of a 2 tesla magnetic field, the BCE for application of 1 kbar of pressure and the eCE for the application of a compressive biaxial strain of 0.05\%. There is fair agreement in both the magnitudes and the signs of these values between theory and experiment, e.g.\ we find that the MCE is inverse and the mechanocaloric (compressive) effects are conventional.

\section{The magnetocaloric effect in the heavy rare earth elements}
\label{HRE}

\begin{figure}[t]
\includegraphics[clip,scale=0.435]{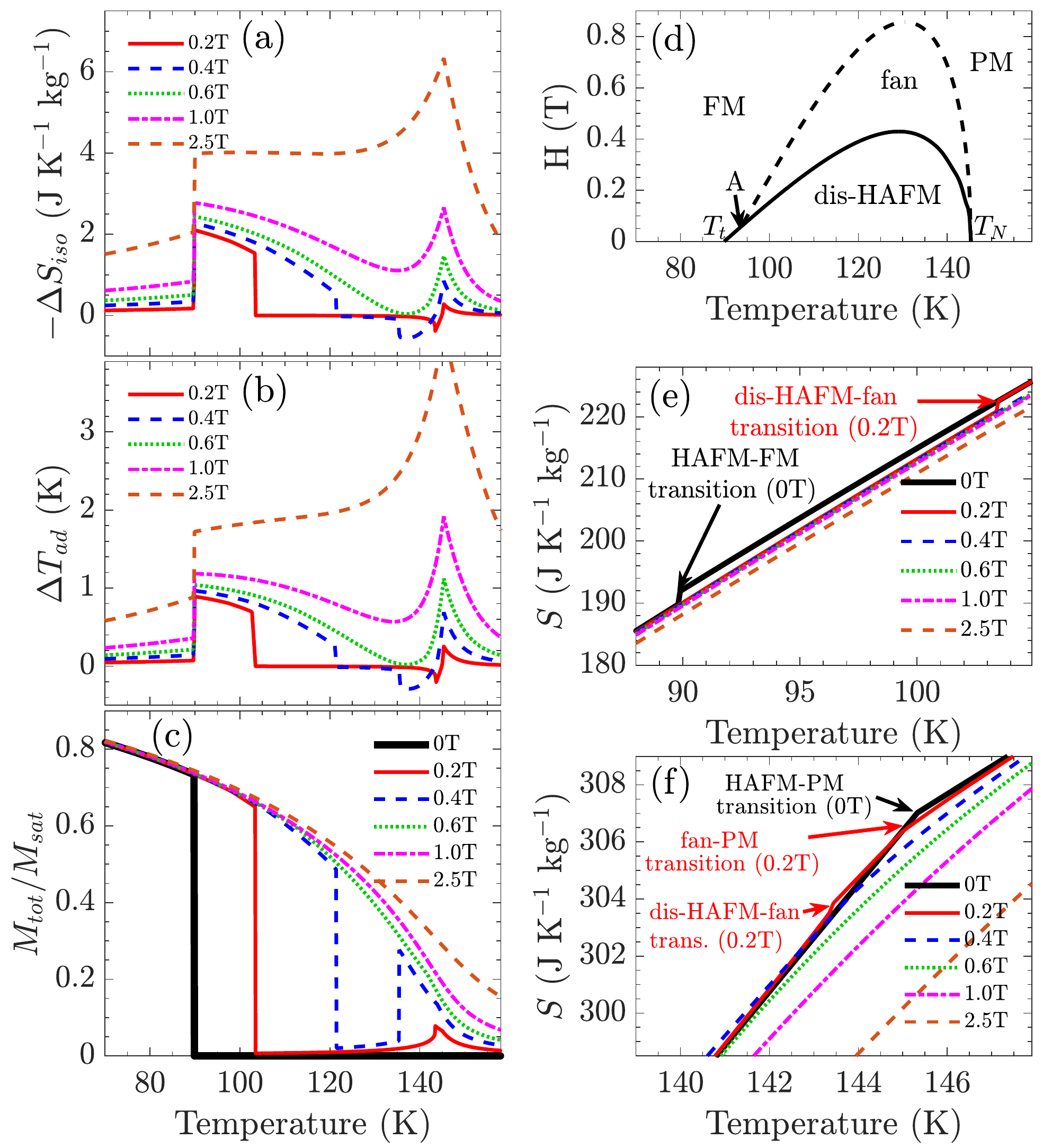}
\caption{(a) Isothermal entropy change and (b) adiabatic temperature change obtained for Dy as functions of temperature produced by the application of a magnetic field from $\textbf{H}=\textbf{0}$ to increasing values up to 2.5 Tesla. (c) The total magnetization along the ferromagnetic component, which is also the direction of the magnetic field, normalized to the saturation value at zero temperature $M_{sat}$. (d) The \textit{ab-initio} temperature-magnetic field phase diagram of Dy as obtained in Ref.~\cite{PhysRevLett.118.197202}. Discontinuous(continuous) black lines indicate first-(second-)order magnetic phase transitions. Panels (e) and (f) show the entropy against temperature around $T_t$ and $T_N$, respectively, for the same increasing values of the applied field.}
\label{DyFig}
\end{figure}

The magnetism of the heavy rare earth (HRE) elements, from gadolinium to lutetium, is composed of a varied range of magnetic phase transitions between complicated, incommensurate AFM and FM states, triggered by both temperature changes and the application of a magnetic field~\cite{Mackintosh1}. Although the HREs share a common valence electronic structure their magnetic phases can be very different. In Ref.~\cite{PhysRevLett.118.197202} we demonstrated that a major part of this complex magnetic phase landscape arises from the response of the valence electrons, via a Fermi surface topological change, to temperature-dependent f-electron magnetic moment order and its link to quartic order magnetic correlations, $f^{(4)}$ (see Eq.\ \ref{Eq2}). A generic temperature-magnetic field phase diagram of a heavy
lanthanide was obtained by minimizing the Gibbs free energy in Eq.\ (\ref{Eq1}), as shown in Fig. \ref{DyFig}(d) for Dy and experimentally observed for Tb~\cite{0953-8984-26-6-066001,0305-4608-14-12-027,PhysRevB.84.132401,PSSA:PSSA2211140164}, Dy~\cite{HERZ1978273,PhysRevB.71.184410,PhysRevB.91.014404}, and Ho~\cite{0953-8984-27-14-146002,doi:10.1063/1.367685}. In the absence of a magnetic field the PM state of these lanthanide metals becomes unstable to the formation of a helical antiferromagnetic (HAFM) ordering through a second-order transition at $T_N$. Lowering the temperature sufficiently triggers a first-order transition to a FM phase at $T_t$ which is stable down to $T=0$K. Whilst the application of a magnetic field $\textbf{H}$ only spin-polarizes further the FM state below $T_t$, it triggers complex magnetic stabilities when applied between $T_t$ and $T_N$ where the helical structure is stable. For increasing values of $\textbf{H}$ the HAFM phase firstly distorts and then discontinuously transforms to a fan magnetic phase~\cite{Mackintosh1}. Further increase of the applied magnetic field continuously stabilizes a FM state.

\begin{figure*}[t]
\centering
\includegraphics[clip,scale=0.42]{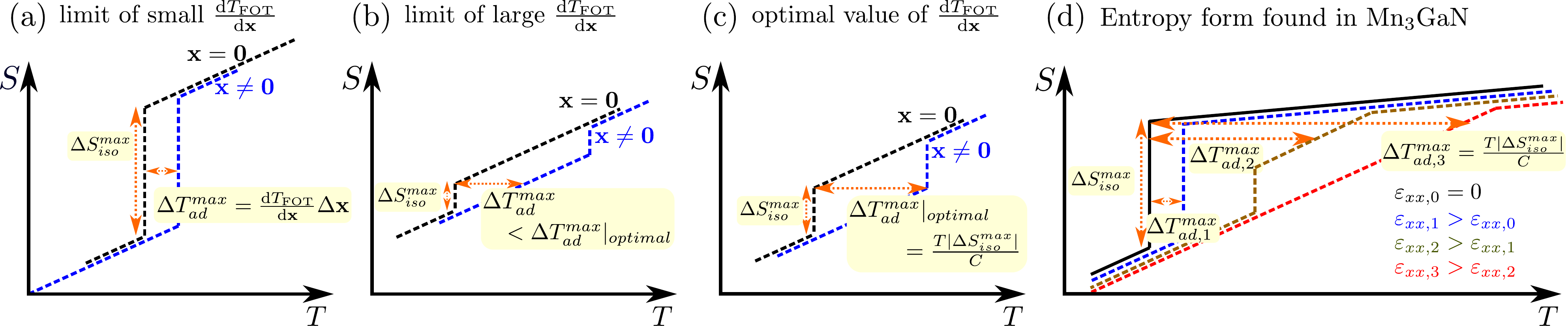}
\caption{Typical temperature-dependent entropy functions of magnetic materials showing a single first-order magnetic phase transition in the limit of (a) small and (b) large values of $\frac{\text{d} T_\text{FOT}}{\text{d}\textbf{x}}$. Panel (c) shows the function for the optimal value of $\frac{\text{d} T_\text{FOT}}{\text{d}\textbf{x}}$ maximizing the available adiabatic temperature change, $\Delta T_{ad}^{max}|_{optimal}$~\cite{SANDEMAN2012566}. (d) Schematics of the entropy form found in Mn$_3$GaN, which shows both first-order and second-order transitions, for increasing values of applied tensile strain.}
\label{FIGLimits}
\end{figure*}

Such a complex magnetic phase diagram contains both second- and first-order magnetic phase transitions which vary differently as a magnetic field is applied. Here we focus on the associated magnetocaloric effect (MCE) which shows a complicated dependence on temperature spanning the large interval between $T_N$ and $T_t$. In panels (a) and (b) of Fig.\ \ref{DyFig} we show the isothermal entropy change, $\Delta S_{iso}$, and adiabatic temperature change, $\Delta T_{ad}$, obtained for Dy, which are in very good qualitative and quantitative agreement with experiment~\cite{PhysRevB.71.184410}. To describe the lattice vibrations we used a Debye temperature of $\theta_D=180$K~\cite{PhysRev.174.504,PALMER1970143,Bodryakov1999}. Values of $\Delta S_{iso}$ and $\Delta T_{ad}$  obtained for Gd, which only shows a single PM-FM second-order transition, also match experiment very well~\cite{0953-8984-26-27-274210}.
In the case of Dy we observe that for smaller values of the magnetic field (below 0.2T) the magnetocaloric response is much larger around $T_t$ compared to $T_N$, as these are first- and second-order transitions respectively. Interestingly, the MCE is conventional at $T_t$ and above $T_N$, but inverse below $T_N$ (see Fig.\ \ref{DyFig}(a,b)). Such a behavior can be traced back to the temperature dependence of the total magnetization, shown in Fig.\ \ref{DyFig}(c). For negative or positive values of $\frac{\partial M_{tot}}{\partial T}$ at small applied fields the MCE is conventional ($\Delta S_{iso}<0$) or inverse ($\Delta S_{iso}>0$), respectively. This directly follows from the fundamental relation $\Delta S_{iso}=\int_{0}^{\textbf{H}}\frac{T}{C}\frac{\partial M_{tot}}{\partial T}d\textbf{H}$~\cite{Tishin}, where $C$ is the heat capacity at constant applied field. The only qualitatively behavior not captured by our theory in comparison with experiment is observed immediately above $T_t$, for which the gradient of the MCE with temperature is negative, i.e.\ $\frac{-\partial\Delta S_{iso}}{\partial T}<0$ and $\frac{\partial\Delta T_{ad}}{\partial T}<0$, instead of positive (see Fig.\ 22 of Ref.~\cite{PhysRevB.71.184410}). This discrepancy is caused by a consistently incorrect sign of $\frac{\partial M_{tot}}{\partial T}$ in the vicinity above $T_t$ for low values of $\textbf{H}$, which is positive in our calculations and negative in experiment. After closely comparing the experimental and theoretical magnetic phase diagrams of Dy~\cite{PhysRevB.71.184410}, we consider that such a minor detail could be related to possible spin-flops or similar magnetic transitions not captured by our theory and linked to subtle magnetization changes.

Inspecting Fig.\ \ref{DyFig}(b) reveals that for large magnetic fields above 0.6T the adiabatic temperature change around the second-order transition $T_N$ becomes bigger than the one observed around the first-order $T_t$, as found in experiment (see Figs.\ 23 and 26 of Ref.~\cite{PhysRevB.71.184410}). This observation can be understood by considering the Clausius-Clapeyron equation~\cite{Tishin,PhysRevLett.100.125901,SANDEMAN2012566}
\begin{equation}
\Delta S_{iso} \propto \left[\frac{\text{d}T_\text{FOT}}{\text{d}\textbf{x}}\right]^{-1}
,
\label{CCeq}
\end{equation}
where $T_\text{FOT}$ is the temperature at which the first-order transition occurs and $\textbf{x}$ is the applied field for the conjugate displacement, i.e.\ a magnetic field $\textbf{H}$ or a mechanical stress $\sigma_{\alpha\beta}$ for magnetization and strain variations, respectively.
Typically, the temperature-dependent entropy behavior observed around first-order magnetic phase transitions is as shown in Fig.\ \ref{FIGLimits}~\cite{SANDEMAN2012566}. Panels (a) and (b) show the limits in which $\frac{\text{d}T_\text{FOT}}{\text{d}\textbf{x}}$ is very small and very large, respectively. Case (a) allows for wide values of $\Delta S_{iso}$ in exchange of small adiabatic temperature changes $\Delta T_{ad}$. Increasing $\frac{\text{d}T_\text{FOT}}{\text{d}\textbf{x}}$ to very high values reduces both $\Delta T_{ad}$ and $\Delta S_{iso}$ (panel (b)), which is in fact the situation observed for Dy at $T_t$ as shown in Fig.\ \ref{DyFig}(e). Note that $\Delta S_{iso}$ is fairly small despite the first-order transition.

The optimal dependence giving the largest possible value of $\Delta T_{ad}$ is obtained at an intermediate magnitude of $\frac{\text{d}T_\text{FOT}}{\text{d}\textbf{x}}$, shown in Fig.\ \ref{FIGLimits}(c), which provides $\Delta T_{ad}^{max}|_{optimal}=\frac{T\Delta S_{iso}^{max}}{C}$ for a substantial value of $\Delta S_{iso}^{max}$ at the transition. However, here $\Delta S_{iso}^{max}$ is far of being the largest possible isothermal entropy change available in the material. If a larger value of the isothermal entropy change is desired it directly follows from Eq.\ (\ref{CCeq}) that $\Delta T_{ad}$ is reduced.
In section \ref{Antiperovs} we discuss how a qualitatively different behavior of the entropy as a function of temperature, found in the magnetically frustrated Mn$_3$GaN antiperovskite material~\cite{PhysRevB.95.184438}, can be used to achieve the maximum possible adiabatic temperature change $\Delta T_{ad}^{max}|_{optimal}$~\cite{SANDEMAN2012566,doi:10.1063/1.3607279,doi:10.1063/1.3309769}, while $\Delta S_{iso}^{max}$ itself is not constrained by the magnitude of $\Delta T_{ad}^{max}|_{optimal}$. Large values for both caloric quantities can then ensue. This desirable outcome arises from a combination of first- and second-order magnetic phase transitions produced by the application of strain and its effect on the geometrically frustrated magnetic interactions between Mn atoms. We expand on such a temperature-dependence of the entropy and compare it with the well studied limits and optimal forms of $S(T)$, shown in Fig.\ \ref{FIGLimits}, and how the presence of a tricritical point enables hysteresis effects to be removed from half of an elastocaloric cooling cycle.

\section{Frustrated magnetism and optimal forms of the entropy}
\label{Antiperovs}

\begin{figure}[h!]
\centering
\includegraphics[clip,scale=0.135]{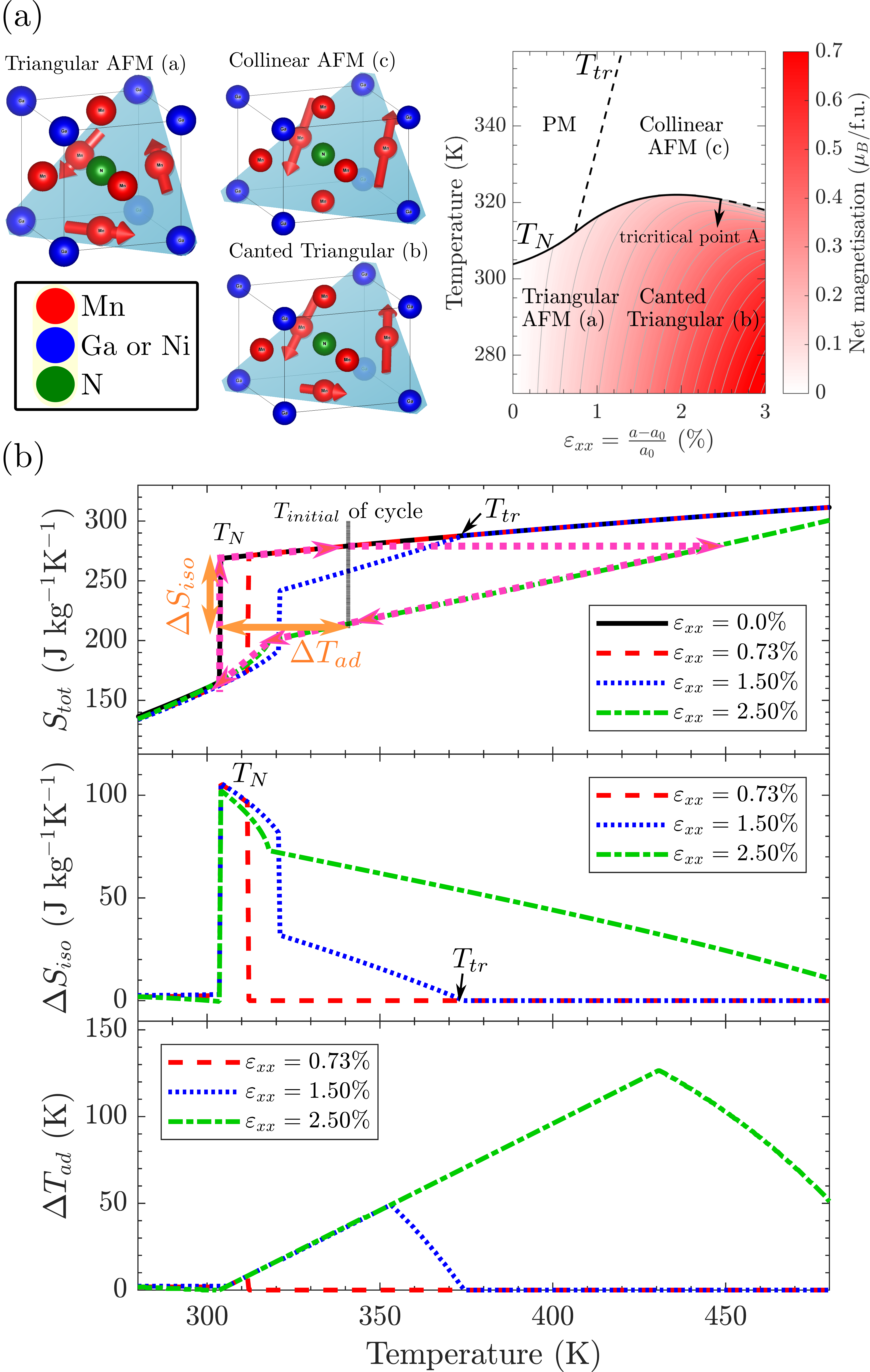}
\caption{(a) Temperature-strain magnetic phase diagram of Mn$_3$GaN, as obtained in Ref.~\cite{PhysRevB.95.184438}, for tensile values of biaxial strain $\varepsilon_{xx}$. The most stable AFM phases, triangular, canted triangular, and collinear, are also shown.
Continuous(dashed) black lines mark first-(second-) order magnetic phase transitions.
(b) Entropy, isothermal entropy change $\Delta S_{iso}$, and adiabatic temperature change $\Delta T_{ad}$ in Mn$_3$GaN as functions of temperature and for increasing values of biaxial strain from zero to $\varepsilon_{xx}=2.5$\%, beyond the tricritical point. $\Delta T_{ad}$ is given for its magnitude obtained when strain is released, i.e.\ $S(T,\varepsilon_{xx}\neq\textbf{0})=S(T+\Delta T_{ad},\varepsilon_{xx}=\textbf{0})$.}
\label{FIGMPD}
\end{figure}

The magnetism of systems presenting strong spin-lattice coupling can be manipulated by the application and removal of mechanical stress. 
Materials in which magnetic frustration plays an important role show particularly dramatic responses when they are strained. Changing their crystal symmetry and relative interatomic distances via stress application can have a substantial impact on the geometrically frustrated magnetic interactions, which in turn strongly change the stability of frustrated and non-frustrated magnetic phases. This is the situation observed in the Mn-based antiperovskite nitride systems Mn$_3$GaN and Mn$_3$NiN~\cite{PhysRevB.95.184438,doi:10.1002/adfm.201902502,PhysRevX.8.041035}. The unstrained form of these two materials show a first-order PM-to-triangular AFM phase transition at $T_N$~\cite{LB1981,Matsunami1}, shown in Fig.\ \ref{FIGMPD}(a), which arises from the frustrated antiferromagnetic interactions between nearest neighbor Mn atoms. The figure shows the theoretically predicted temperature-biaxial strain magnetic phase diagram of Mn$_3$GaN, obtained using the theory presented here~\cite{PhysRevB.95.184438}, which is qualitatively similar to the experimental diagram recently found for Mn$_3$NiN~\cite{doi:10.1002/adfm.201902502}. As tensile biaxial strain is increased and the $c/a$ ratio reduces, the triangular AFM phase distorts to form a canted triangular AFM state at low temperature. Interestingly, at higher temperatures the PM state becomes unstable to the formation of a collinear AFM phase at $T_{tr}$, which exhibits a very prominent shift upwards as biaxial strain increases.
This collinear AFM state shows itself as the most magnetically frustrated phase in the diagram since one of its Mn sites is fully frustrated and a zero net magnetizations occurs, i.e.\ the corresponding orientational average of the magnetic moment vanishes, $\textbf{m}_n=\langle\hat{\bf{e}}_n\rangle=\textbf{0}$.
The transitions between all these magnetic phases have different first- and second-order character, as denoted in the diagram, and the associated entropy change at $T_N$ is giant~\cite{Matsunami1}. This makes Mn-based antiperovskite materials very interesting for the study of caloric effects emerging from the change of magnetism driven by mechanical stimuli, i.e.\ a mechanocaloric effect.

In Fig.\ \ref{FIGMPD}(b) we show the calculated temperature dependence of the entropy, schematically shown in Fig.\ \ref{FIGLimits}(d), the isothermal entropy change and the adiabatic temperature change obtained for Mn$_3$GaN on releasing the external stress as $\varepsilon_{xx}\neq 0 \rightarrow \varepsilon_{xx}=0$. 
At small strains the entropy's temperature-dependent behavior corresponds to the small $\frac{\text{d}T_N}{\text{d}\sigma}$ limit, where $\sigma$ is the absolute value of the biaxial tensile stress applied. This situation gives substantial values of $\Delta S_{iso}$ (see Eq.\ (\ref{CCeq})) but modest $\Delta T_{ad}$ changes. However, for larger strains $S(T)$ strongly changes its behavior to resemble the temperature-dependence observed in the limit of large $\frac{\text{d}T_N}{\text{d}\sigma}$. This is a direct consequence of its approaching the tricitical point A shown in the magnetic phase diagram (Fig.\ \ref{FIGMPD}(a)). At larger strains from this point the first-order phase transition between the collinear AFM and canted triangular AFM phases becomes second-order and co-exists together with a PM-collinear AFM second-order phase transition occurring at high temperature. The changing shape of $S(T)$ with stress application from one limiting case to the other gives rise to a somewhat triangular contour, in contrast with the rectangular one typically observed shown in Fig.\ \ref{FIGLimits}(a-c). This automatically produces a maximum adiabatic temperature change equal to $\Delta T_{ad}^{max}=\frac{T\Delta S_{iso}^{max}}{C}$, provided that large enough tensile strains are attained.
This is the largest possible value allowed for an arbitrary magnitude of $\Delta S_{iso}$ associated with the first-order magnetic phase transition at $\varepsilon_{xx}=0$, and also equal to the optimal value for materials with a single discontinuous transition~\cite{SANDEMAN2012566}, see Fig.\ \ref{FIGLimits}.
The combination of first- and second-order magnetic phase transitions, and the presence of the tricritical point A, produces a changing entropy shape against applied field which maximizes both the isothermal entropy change and adiabatic temperature change simultaneously. A key aspect in the origin of this optimal behavior is how the instability of the PM phase to the formation of a magnetic state changes from a strong first-order with small $\frac{\text{d}T_N}{\text{d}\sigma}$ (small $\varepsilon_{xx}$) to a second-order with large $\frac{\text{d}T_N}{\text{d}\sigma}$ (large $\varepsilon_{xx}$). Here this occurs owing to the highly responsive frustrated magnetic interactions between Mn atoms to strain application~\cite{PhysRevB.95.184438}.

At $\varepsilon_{xx}=2.5$\% a temperature span of roughly 100K above $T_N\approx 300$K is obtained where both $\Delta S_{iso}$ and $\Delta T_{ad}$ are very large, as shown in Fig.\ \ref{FIGMPD}(b). We highlight that the experimental magnetic phase diagram recently obtained for Mn$_3$NiN~\cite{doi:10.1002/adfm.201902502} shows that 0.1\% biaxial strain is enough to stabilize the collinear AFM phase at very high temperature, which is an order of magnitude smaller than our predictions. This indicates that getting close to the vicinity of the tricritical point would require the production of around 0.25\% strains, which should be experimentally feasible. Finally, we point out that mechanocaloric cooling cycles using strains above the tricritical point, for example $\varepsilon_{xx}=2.5$\% (expected to be lower in experiment), will strongly benefit from the removal of hysteresis effects on half of an elastocaloric caloric cycle. As shown for the proposed cooling cycle indicated by dotted magenta lines in the top plot of Fig.\ \ref{FIGMPD}(b), a first-order transition is crossed only a single time every cycle owing to the second-order behavior at $\varepsilon_{xx}=2.5\%$.

\section{Conclusions}
\label{Conc}
New multicaloric effects and cooling cycles in which $\Delta S_{iso}$ and $\Delta T_{ad}$ quantities are large and hysteresis losses small are highly desirable for the ongoing development of solid state cooling technologies~\cite{doi:10.1002/ente.201800264}. At a fundamental level the effects stem from the complexity of the interacting electronic glue in materials, how it responds and affects magnetic order, for example, and links it with other physical properties. In this article we have set out a case for {\it ab initio} computational materials modelling to have a key role in harnessing this physics, providing insight, sometimes being predictive and capturing trends well. We have presented theory results and compared them with experiment for caloric effects in three materials, the magnetic phase diagrams of which we had previously modelled with our DFT-DLM theory. All three materials possess both first- and second-order magnetic phase transitions as a function of $T$ and applied stimulus and we have shown how the sign and size of the caloric  $\Delta S_{iso}$ and $\Delta T_{ad}$ quantities vary in the vicinity of these transitions. The link to the variation of transition temperatures with applied stimulus has also been studied. We have further exploited the rich phase diagram a Mn-rich antiperovskite nitride to propose an example of a new sort of cooling cycle in which a major source of hysteresis effects has been removed. This demonstration might motivate further exploration of the solid state cooling potential of Mn-rich materials which harbor frustrated magnetic interactions.

\section{Acknowledgements}
The work was supported by EPSRC (UK) grants EP/J06750/1 and EP/M028941/1. E.\ M.-T. acknowledges funding from the priority programme SPP1599 ”Ferroic Cooling” (Grant No. HI1300/6-2).

\bibliography{./bibliography.bib}
\end{document}